\documentclass[twocolumn,english]{aastex6}
\usepackage[latin9]{inputenc}
\usepackage{xcolor}
\usepackage{array}
\usepackage{graphicx}
\usepackage{setspace}
\PassOptionsToPackage{normalem}{ulem}
\usepackage{ulem}

\makeatletter

\providecommand{\tabularnewline}{\\}
\providecolor{lyxadded}{rgb}{0,0,1}
\providecolor{lyxdeleted}{rgb}{1,0,0}

\let\jnl@style=\rm
\def\ref@jnl#1{{\jnl@style#1}}

\def\aj{\ref@jnl{AJ}}                   
\def\actaa{\ref@jnl{Acta Astron.}}      
\def\araa{\ref@jnl{ARA\&A}}             
\def\apj{\ref@jnl{ApJ}}                 
\def\apjl{\ref@jnl{ApJ}}                
\def\apjs{\ref@jnl{ApJS}}               
\def\ao{\ref@jnl{Appl.~Opt.}}           
\def\apss{\ref@jnl{Ap\&SS}}             
\def\aap{\ref@jnl{A\&A}}                
\def\aapr{\ref@jnl{A\&A~Rev.}}          
\def\aaps{\ref@jnl{A\&AS}}              
\def\azh{\ref@jnl{AZh}}                 
\def\baas{\ref@jnl{BAAS}}               
\def\bac{\ref@jnl{Bull. astr. Inst. Czechosl.}}
\def\caa{\ref@jnl{Chinese Astron. Astrophys.}}
\def\cjaa{\ref@jnl{Chinese J. Astron. Astrophys.}}
\def\icarus{\ref@jnl{Icarus}}           
\def\jcap{\ref@jnl{J. Cosmology Astropart. Phys.}}
\def\jrasc{\ref@jnl{JRASC}}             
\def\memras{\ref@jnl{MmRAS}}            
\def\mnras{\ref@jnl{MNRAS}}             
\def\na{\ref@jnl{New A}}                
\def\nar{\ref@jnl{New A Rev.}}          
\def\pra{\ref@jnl{Phys.~Rev.~A}}        
\def\prb{\ref@jnl{Phys.~Rev.~B}}        
\def\prc{\ref@jnl{Phys.~Rev.~C}}        
\def\prd{\ref@jnl{Phys.~Rev.~D}}        
\def\pre{\ref@jnl{Phys.~Rev.~E}}        
\def\prl{\ref@jnl{Phys.~Rev.~Lett.}}    
\def\pasa{\ref@jnl{PASA}}               
\def\pasp{\ref@jnl{PASP}}               
\def\pasj{\ref@jnl{PASJ}}               
\def\rmxaa{\ref@jnl{Rev. Mexicana Astron. Astrofis.}}%
\def\qjras{\ref@jnl{QJRAS}}             
\def\skytel{\ref@jnl{S\&T}}             
\def\solphys{\ref@jnl{Sol.~Phys.}}      
\def\sovast{\ref@jnl{Soviet~Ast.}}      
\def\ssr{\ref@jnl{Space~Sci.~Rev.}}     
\def\zap{\ref@jnl{ZAp}}                 
\def\nat{\ref@jnl{Nature}}              
\def\iaucirc{\ref@jnl{IAU~Circ.}}       
\def\aplett{\ref@jnl{Astrophys.~Lett.}} 
\def\apspr{\ref@jnl{Astrophys.~Space~Phys.~Res.}}
\def\bain{\ref@jnl{Bull.~Astron.~Inst.~Netherlands}} 
\def\fcp{\ref@jnl{Fund.~Cosmic~Phys.}}  
\def\gca{\ref@jnl{Geochim.~Cosmochim.~Acta}}   
\def\grl{\ref@jnl{Geophys.~Res.~Lett.}} 
\def\jcp{\ref@jnl{J.~Chem.~Phys.}}      
\def\jgr{\ref@jnl{J.~Geophys.~Res.}}    
\def\jqsrt{\ref@jnl{J.~Quant.~Spec.~Radiat.~Transf.}}
\def\memsai{\ref@jnl{Mem.~Soc.~Astron.~Italiana}}
\def\nphysa{\ref@jnl{Nucl.~Phys.~A}}   
\def\physrep{\ref@jnl{Phys.~Rep.}}   
\def\physscr{\ref@jnl{Phys.~Scr}}   
\def\planss{\ref@jnl{Planet.~Space~Sci.}}   
\def\procspie{\ref@jnl{Proc.~SPIE}}   

\makeatother

\usepackage{babel}
\begin{document}

\title{Spiral arms, warping, and clumps formation\\
 in the Galactic center young stellar disk}
\author{Hagai B. Perets$^{1}$, Alessandra Mastrobuono-Battisti$^{1,2}$,
Yohai Meiron$^{3}$ and Alessia Gualandris$^{4}$ }
\affil{$^{1}$Physics department, Technion - Israel institute of Technology,
Haifa, Israel, 3200002\\
$^{2}$Max Planck Institute for Astronomy, Königstuhl 1769117, Heidelberg,
Germany\\
$^{3}$Eotvos University, Budapest, Pazmany Peter setany 1/A, Institute
of Physics, Hungary, 1117\\
$^{4}$University of Surrey, Physics department, 388 Stag Hill, Guildford
GU2 7XH, UK}
\begin{abstract}
The Galactic center of the Milky-Way harbors a massive black hole
(BH) orbited by a diverse population of young and old stars. A significant
fraction of the youngest stars ($\sim4-7$ Myr) reside in a thin stellar
disk with puzzling properties; the disk appears to be warped, shows
asymmetries, and contains one or more clumpy structures (e.g. IRS
13). Models explaining the clumping invoked the existence of an intermediate
mass BH of $10^{3}-10^{4}$ M$_{\odot}$, but no kinematic evidence
for such a BH has been found. Here we use extended $N$-body simulations
and hybrid self-consistent field method models to show that naturally
formed residual temporal asphericity of the hosting nuclear star cluster
gives rise to torques on the disk, which lead to changes in its orientation
over time, and to recurrent formation and dissolution of single spiral
arm ($m=1$ modes) structures. The changing orientation leads to a
flapping-like behavior of the disk and to the formation of a warped
disk structure. The spiral arms may explain the over-densities in
the disk (clumping) and its observed asymmetry, without invoking the
existence of an intermediate mass BH. The spiral arms are also important
for the overall disk evolution, and can be used to constrain the structure
and composition of the nuclear stellar cluster. 
\end{abstract}
\section{Introduction}
Based on detailed kinematic data of stars in the Galactic nucleus,
a massive black hole( MBH; measured at $\sim4.3\times10^{6}$ solar
masses, M$_{\odot}$), is inferred to exist in its center \citep{gen+97,ghe+98,gen+10}.
Similar data have been used to reveal the existence of a disk of young,
4-7 Myrs massive stars orbiting the MBH \citep{lev+03,pau+06,lu+09,bar+10,gen+10},
embedded in a massive few-pc sized, Gyrs old, nuclear stellar cluster
(NSC). Extensive studies of the disk over the last decade have characterized
its unique structure and composition. The disk appears to be warped
\citep{gen+10}, and a significant over-density (clump) was identified
in the disk (the IRS 13 structure; \citealt{Mai+04})

The existence of a very young stellar disk close to the massive black
hole (MBH) in the Galactic center (GC) has been suggested to originate
from the infall of a gas cloud forming a gaseous disk which then fragments
and gives rise to the observed stellar disk \citep{lev+03}. Detailed
hydrodynamical simulations of nuclear-disk star-formation have been
able to produce stellar disks with similar global properties as those
observed in the GC in terms of mass, stellar mass function and radial
density profile \citep{Nay+07,Bon+08}. However, any initial inhomogeneities
in the disk were found to be rapidly erased after a few orbits, leaving
behind a smooth disk. The later evolution of the disk is affected
by two-body relaxation processes due to stars in the disk and in the
host NSC. Such processes kinematically heat the disk (i.e. increase
the velocity dispersion of the disk stars), leading to higher eccentricities
and disk thickening \citep{ale+06,Mik+17,Nao+18}, but are not expected
to produce any of the aforementioned peculiar kinetic features (clumping,
warping, asymmetries). The clump has been suggested to result from
an IMBH residing in the disk; however, detailed kinematic studies
of IRS13 ruled out the existence of such object \citep{Mai+04,gen+10}.
The origins of the clump, as well as the disk warping, therefore remained
open questions. Here we show that \emph{collective} processes operate
in such an environment and give rise to such properties, potentially
explaining their origin in the GC disk. 

In order to study the evolution of the nuclear disk, we simulated
the evolution of both the stellar disk as well as the background stellar
population of the nuclear stellar cluster (NSC) in which the disk
is embedded, using full (i.e. direct) \emph{$N$}-body simulations.
Due to the computational cost of such simulations, they can not follow
the large number of stars in the NSC, and we therefore complemented
these simulations with a hybrid self-consistent field (SCF) modeling
\citep{Mei+14} to validate and extend our results, for a realistic
NSC accountinf for all the stars. Together our models show that collective
processes robustly determine the evolution of the disk and may give
rise to its peculiar properties.

We begin by describing the disk models we considered and the methods/codes
used to simulate the disk evolution (section \ref{sec:simulations}).
We then decribe our results (section 2), followed by the discussion
(section 3) and summary (section 4) 

\section{Simulation models}
\label{sec:simulations}
\subsection{Disk and cusp models }
Observations show the existence of a stellar cusp of typically low-mass,
old, stellar population, in which the disk is embedded; theoretical
studies suggest that it includes a significant population of compact
objects, most importantly stellar black holes (SBHs; e.g. \citep{mir+00,hop+06b,fre+06,Aha+16}).
Our initial conditions were chosen as to best realize the initial
properties of the disk and the NSC, given their observed, inferred
and modeled characteristics. The considered a disk extending between
$0.03$ and $0.3$pc (we also considered $0.01$pc inner cutoffs with
no qualitative differences observed), with a power-law surface density
profile ($\propto r^{-\gamma};$ $1.5\le\gamma\le2$) and a total
disk mass of a few$\times10^{3}$ M$_{\odot}$up to $\sim10^{4}$
M$_{\odot}$ \citep{gen+10}. We assumed that the initial disk was
thin following its formation \citep{Nay+07}, taking the initial disk
height to be $0.025$pc (5$\%$ of the disk length). The masses of
disk stars were sampled from a Salpeter mass function, between $0.6$
and $60$M$_{\odot}$ (though as we later show, the masses of the
disk stars play a negligible role). The MBH position was kept constant
as in a realistic NSC, only little Brownian motion of the MBH is expected
\citep{mer04a}. Nevertheless we also modeled cases with freely moving
MBH (not discussed here; which showed qualitatively similar behaviors). 

Following the evolution of all stars in a realistic NSC is computationally
prohibitive, and we therefore first modeled only the population of
SBHs in the spherical cusp, which is the dynamically dominant population
in this region. The rest of the NSC mass was initially modeled as
an external smooth potential, not following the evolution of the individual
stars in the NSC. As we discuss later on, we also used a hybrid model
consisting of a self-consistent field (SCF) method to model the NSC
coupled with an $N$-body code for modeling the disk itself \citep{Mei+14}.
The hybrid modeling allowed us to explore a more realistic NSC (on
the expense of neglecting 2-body relaxation processes in the NSC).
We have explored several modifications of the initial conditions and
models (as described in Table 1). We explored the cases where the
disk is evolved in an NSC modeled as a smooth external potential (or
with no NSC at all), i.e. not including the stochastic effects introduced
by individual stars; we considered the effects of changing the SBHs
masses (while keeping the total mass of the SBH population constant);
and we explored the case when the disk stars are treated as test-particles
as to exclude disk self-gravity effects. 
The full details of all the models studies can be found in Table \ref{tab:models},
where we also list the specific codes used (see below). 

\begin{figure*}
\includegraphics{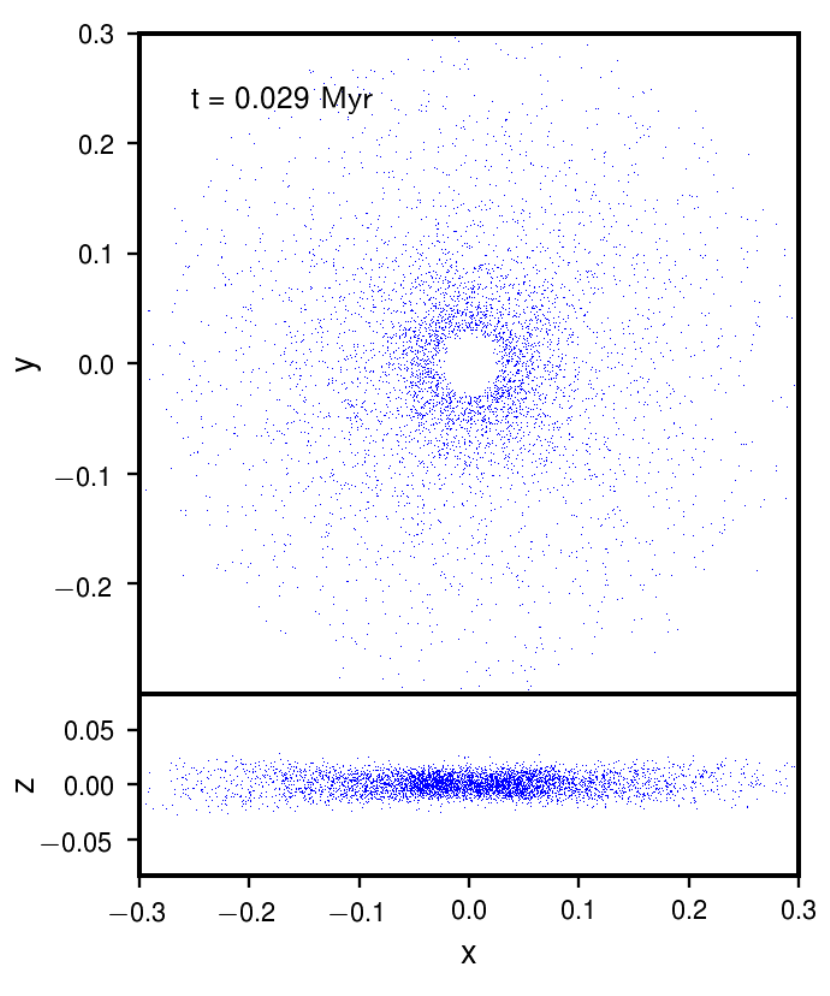}\includegraphics{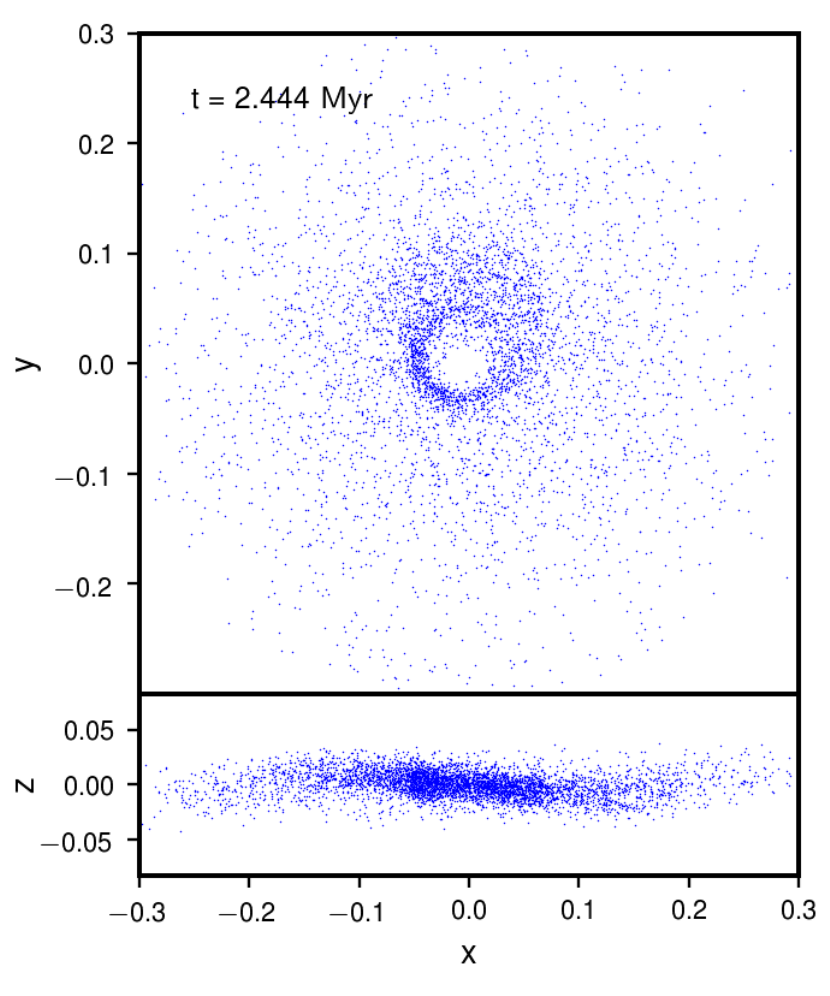}

\includegraphics{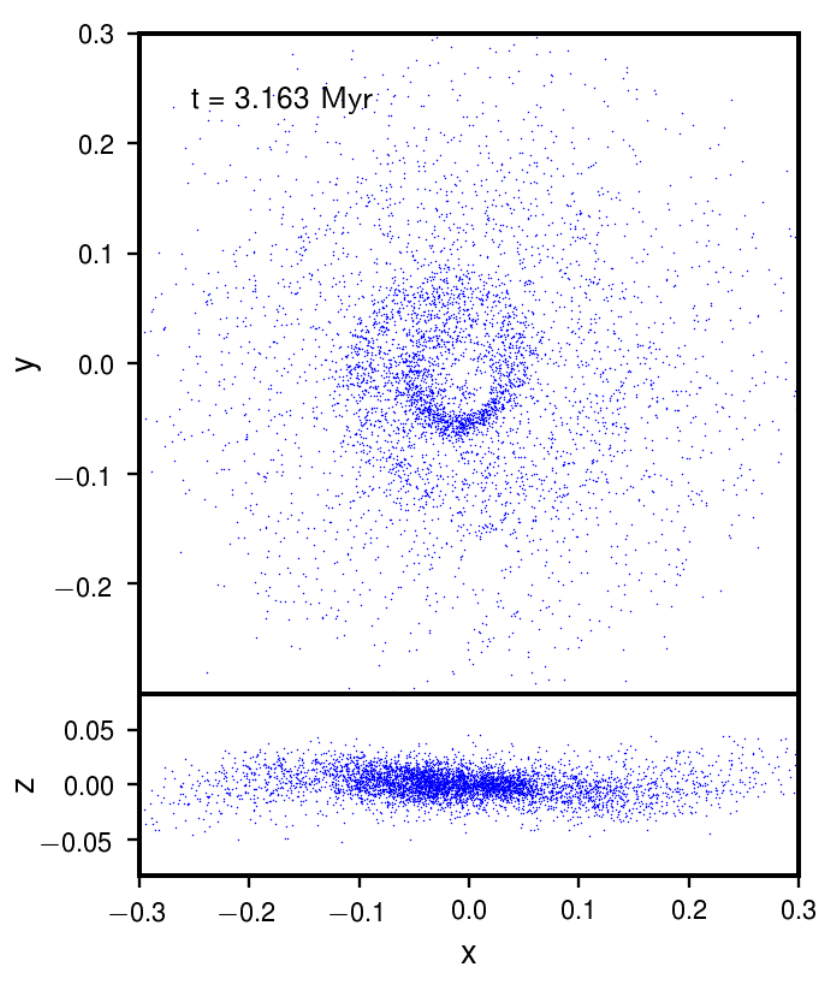}\includegraphics{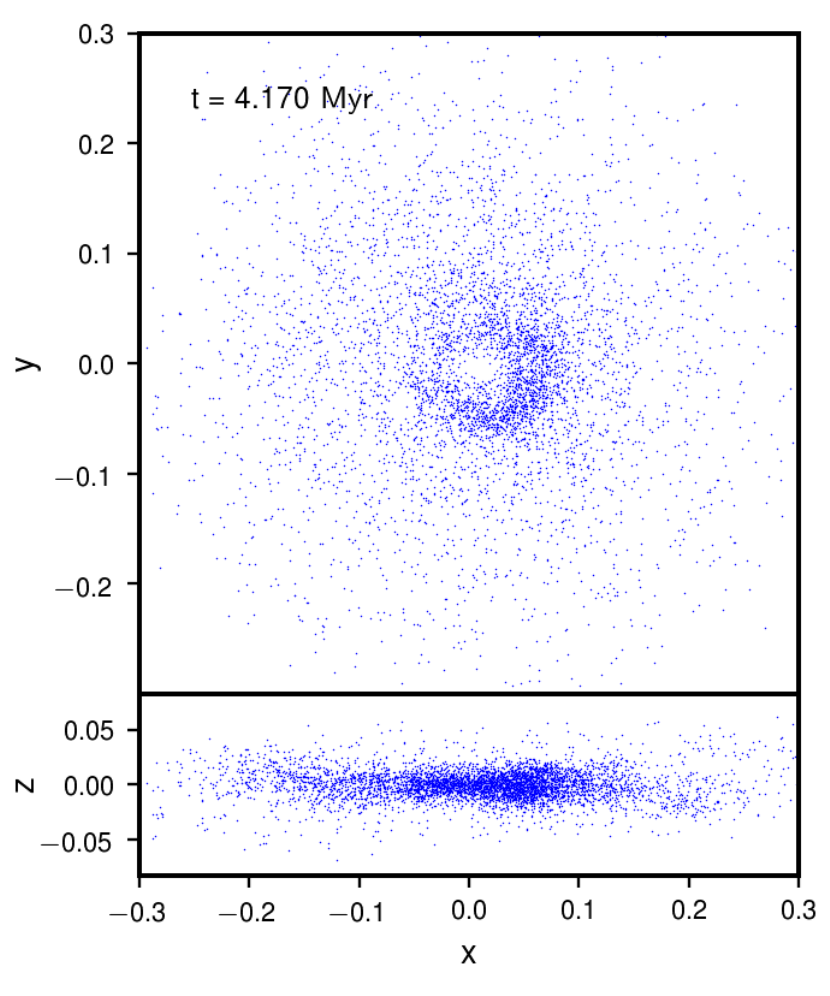}\caption{\label{fig:disk-evolution}
The evolution of a nuclear stellar disk embedded in a live NSC over
a few Myrs of evolution (model 9). The NSC contains both stellar and
SBH components, and the system hosts a $4.3\times10^{6}\,{\rm M}_{\odot}$
MBH as observed in the Galactic center. Shown are four edge-on and
face-on snapshot views of the disk at several times. The formation
and evolution of spiral arm structure and warping are clearly seen.
The detailed properties of disk and the hosting NSC are given in Table
\ref{tab:models}. }
\end{figure*}

\begin{table*}
{\footnotesize{}}%
\begin{tabular}{|>{\centering}p{0.2cm}|>{\centering}p{2.5cm}|>{\centering}p{1.7cm}|c|>{\centering}p{1.5cm}|>{\centering}p{1.3cm}||>{\centering}p{3.2cm}|}
\hline 
{\footnotesize{}\#} & {\footnotesize{}Model} & {\footnotesize{}Disk stars} & {\footnotesize{}NSC stars} & {\footnotesize{}NSC SBHs} & Code & Result\tabularnewline
\hline 
 &  & {\footnotesize{}mass function } &  & {\footnotesize{}mass function} &  & \tabularnewline
\hline 
\hline 
{\scriptsize{}1 } & {\footnotesize{}Isolated disk} & {\footnotesize{}Salpeter $0.6-60\,{\rm M_{\odot}}$} & {\footnotesize{}\textemdash{} } & {\footnotesize{}No SBHs} & {\footnotesize{}$N$-body} & {\footnotesize{}stable, smooth non-warped disk}\tabularnewline
\hline 
{\scriptsize{}2 } & {\footnotesize{}Disk+smooth NSC } & {\footnotesize{}Salpeter $0.6-60\,{\rm M_{\odot}}$} & {\footnotesize{}Smooth potential} & {\footnotesize{}No SBHs} & {\footnotesize{}$N$-body} & {\footnotesize{}stable, smooth non-warped disk}\tabularnewline
\hline 
{\scriptsize{}3} & {\footnotesize{}Disk+live SBHs} & {\footnotesize{}Salpeter $0.6-60\,{\rm M_{\odot}}$} & {\footnotesize{}\textemdash{} } & {\footnotesize{}10${\rm M_{\odot}}$} & {\footnotesize{}$N$-body} & {\footnotesize{}unstable, spiral arm, clumping, warping}\tabularnewline
\hline 
{\scriptsize{}4} & {\footnotesize{}Test disk+live SBHs} & {\footnotesize{}Mass-less } & {\footnotesize{}\textemdash{} } & {\footnotesize{}10${\rm M_{\odot}}$} & {\footnotesize{}$N$-body} & {\footnotesize{}unstable, spiral arm, clumping, warping}\tabularnewline
\hline 
{\scriptsize{}5} & {\footnotesize{}Disk+live SBHs +smooth NSC } & {\footnotesize{}Salpeter $0.6-60\,{\rm M_{\odot}}$} & {\footnotesize{}Smooth potential} & {\footnotesize{}10${\rm M_{\odot}}$} & {\footnotesize{}$N$-body} & {\footnotesize{}unstable, spiral arm, clumping, warping}\tabularnewline
\hline 
{\scriptsize{}6} & {\footnotesize{}Disk+live SBHs +smooth NSC } & {\footnotesize{}Salpeter $0.6-60\,{\rm M_{\odot}}$} & {\footnotesize{}Smooth potential} & {\footnotesize{}20${\rm M_{\odot}}$} & {\footnotesize{}$N$-body} & {\footnotesize{}unstable, spiral arm, clumping, warping}\tabularnewline
\hline 
{\scriptsize{}7} & {\footnotesize{}Disk+live SBHs +smooth NSC } & {\footnotesize{}Salpeter $0.6-60\,{\rm M_{\odot}}$} & {\footnotesize{}Smooth potential} & {\footnotesize{}40${\rm M_{\odot}}$} & {\footnotesize{}$N$-body} & {\footnotesize{}unstable, spiral arm, clumping, warping}\tabularnewline
\hline 
{\scriptsize{}8} & {\footnotesize{}Disk+live SBHs +NSC } & {\footnotesize{}Salpeter $0.6-60\,{\rm M_{\odot}}$} & {\footnotesize{}SCF(160k stars; m=4${\rm M_{\odot}})$} & {\footnotesize{}40${\rm M_{\odot}}$} & {\footnotesize{}hybrid} & {\footnotesize{}unstable, spiral arm, clumping, warping}\tabularnewline
\hline 
{\scriptsize{}9} & {\footnotesize{}Disk+live SBHs +NSC} & {\footnotesize{}Salpeter $0.6-60\,{\rm M_{\odot}}$} & {\footnotesize{}SCF(640k stars; m=1${\rm M_{\odot}})$} & {\footnotesize{}10${\rm M_{\odot}}$} & {\footnotesize{}hybrid} & {\footnotesize{}unstable, spiral arm, clumping, warping}\tabularnewline
\hline 
{\scriptsize{}10} & {\footnotesize{}Disk+live SBHs +NSC} & {\footnotesize{}Salpeter $0.6-60\,{\rm M_{\odot}}$} & {\footnotesize{}SCF(640k stars; m=1${\rm M_{\odot}})$} & {\footnotesize{}20${\rm M_{\odot}}$} & {\footnotesize{}hybrid} & {\footnotesize{}unstable, spiral arm, clumping, warping}\tabularnewline
\hline 
{\scriptsize{}11} & {\footnotesize{}Disk+live SBHs +NSC} & {\footnotesize{}Salpeter $0.6-60\,{\rm M_{\odot}}$} & {\footnotesize{}SCF(640k stars; m=1${\rm M_{\odot}})$} & {\footnotesize{}40${\rm M_{\odot}}$} & {\footnotesize{}hybrid} & {\footnotesize{}unstable, spiral arm, clumping, warping}\tabularnewline
\hline 
{\scriptsize{}12} & {\footnotesize{}Disk+NSC} & {\footnotesize{}Salpeter $0.6-60\,{\rm M_{\odot}}$} & {\footnotesize{}SCF(640k stars; m=1${\rm M_{\odot}})$} & {\footnotesize{}No SBHs} & {\footnotesize{}hybrid} & {\footnotesize{}unstable, spiral arm, clumping, warping}\tabularnewline
\hline 
\end{tabular}{\footnotesize{}\caption{\label{tab:models}}
}The models for the disk and nuclear cluster explored with the N-body
($\phi{\rm GRAPE}$ code) and SCF-hybrid (ETICS code) simulations.
Live refers to full $N$-body modeling of the stars; Test particles
refer to the $N$-body modeling of the stars as mass-less test particles.
Hybrid refers to the case where each of the stars is fully represented
on a one to one level, but the overall potential from the NSC stars
is calculated only through the SCF method, while the disk stars are
fully modeled through $N$-body evolution. Smooth potential refers
to spherically symmetric external potential representing the overall
smoothed mass distribution of the stars without the effect of their
discrete representation. 
\end{table*}

\subsection{N-body and hybrid self-self-consistent field method simulations}
In our study we made use of the $\phi$GRAPE \citep{har+07} $N$-body
code (modified to allow the addition of an external potential), using
a GPU-based GRAPE emulation library. In addition, we also employed
the self-developed NBSymple code \citep{Cap+11} to better verify
and check that our results are not numerical artifacts arising from
the use of a specific code. Finally we complemented the $N$-body
simulations using a hybrid code coupling $\phi$GRAPE with the self-consistent
field method ETICS \citep{Mei+14} code. The hybrid code works by
dividing the stellar system into a collisional (C) and collisionless
(S, for soft) parts, in this case C is comprised only of the disk
particles, while S is comprised of the SBHs and NSC stars. The full
force calculation has four components designated CC, CS, SC, and SS.
The first represents the force exerted by collisional (disk) particles
on other collisional particles, the second represents the force exerted
by the collisional particles on the collisionless particles, and so
on. Of those four components, CC is calculated by direct summation
by employing a GRAPE call from an emulation library, while the three
other components are calculated with the SCF method by calling ETICS
subroutines. The forces are then combined in a proper way and returned
to $\phi$GRAPE to perform the time integration in the normal way.
In all cases, the codes produced similar outcomes when similar initial
conditions had been used, and therefore in the following we describe
only qualitatively different simulations done using the $\phi{\rm GRAPE}$
$N$-body code and the hybrid code. The energy conservation in our
simulations was in the range of $10^{-3}-10^{-2}$ after $6$ Myrs
of evolution. We simulated both disks that evolve around the MBH in
isolation without a hosting NSC, as well as disks coupled to a NSC.
The properties of the disks and NSCs explored in our simulations are
described in the following, and a summary of our main models is shown
in Table 1. 

\textbf{Disk:} We explored several models for the evolution of a stellar
disk with global properties comparable to that observed/inferred for
the stellar disk in the GC \citep{gen+10}. All our disks included
5000 stars, which masses are sampled from a Salpeter mass function
in the range $0.6-60$ ${\rm M_{\odot}}$ with a total disk mass of
$\sim9500$ ${\rm M_{\odot}}$. In order to explore the role played
by the self-gravity of the disk we also considered a disk composed
of mass-less test-particles. The latter mass-less disk evolved qualitatively
the same as the realistic massive disk. Our simulated disks extend
between 0.05 and 0.5 pc, comparable to the GC disk. We also considered
a model with a lower inner cutoff (0.03 pc), for which results showed
no qualitative differences from our other models. We considered a
power-law surface density profile for all simulated disk ($\propto r^{-\gamma};$
$1.5\le\gamma\le2$). We assumed the initial disk was thin following
its formation \citep{Nay+07}, taking the initial disk height to be
0.025 pc (5$\%$ of the disk length). 

\textbf{NSC:} Several qualitatively different types of NSC models
were considered. 
\begin{enumerate}
\item \uline{Live-NSCs composed only of SBHs:} These were run with the
$N$-body simulation and included 16k/8k/4k point-mass particles of
10, 20 and 40 ${\rm M}{}_{\odot}$, respectively, where the total
mass in SBHs ($1.6\times10^{5}\,{\rm M_{\odot}}$) had been kept constant.
The SBHs were distributed isotropically around the MBH with a $r^{-2}$
radial density profile between $0.03-0.8$ pc from the MBH (we also
considered lower, $0.01$ pc, and higher $0.04$pc cutoffs, showing
similar qualitative behavior). Such BH cusps are expected to exist
in GC-like nuclei, following the formation and evolution through mass-segregation
processes (e.g. \citealt{hop+06b,fre+06,ale+09,Pre+10,Aha+16}). 
\item \uline{Smooth potential and hybrid live-NSCs:} In order to disentangle
the overall effects of the NSC smooth background potential from the
emergent asphericity effects, arising from the discreteness of the
NSC background stellar population, we also ran models with the NSC
modeled as a smooth spherical gravitational potential. Such models
produced no instabilities. In our $N$-body models we included a smooth
potential for the large stellar NSC, and a live, full $N$-body model
of the SBHs-only cases.
\item \uline{Realistic live-NSCs:} These included both the SBHs (as in
the SBHs-only models), as well as a realistic live stellar NSC, composed
of $6.4\times10^{5}$ ($1\,{\rm M_{\odot}}$stars), distributed isotropically
with a $r^{-7/4}$ density profile over the same $0.03-0.8$ pc range
as the SBHs. We also considered a case of a live NSC with no SBHs,
composed only of the stellar component. Since full $N$-body simulations
of such realistic NSCs are too computationally expensive to run with
direct $N$-body simulations, the realistic NSC models were all run
using our hybrid code.
\end{enumerate}
\section{Results}
Our simulations confirm the results of previous models in the case
of the evolution of an isolated (no NSC) stellar disk (e.g. \citep{ale+06,Sub+14,Mik+17}
and references therein), showing only a steady disk-height increase
over its 6 Myrs evolution, consistent with 2-body relaxation models.
Such models produce a smooth (no clumps or asymmetries), aligned (no
warping) structure which does not change its orientation. Adding a
constant smooth potential to model the NSC shows very similar results.
However, once we include the effects of individual SBHs in the NSC
(a ``live'' NSC) the picture changes dramatically (see Figures 1-2).
The \emph{collective} variable contribution of the stars in the NSC
gives rise to naturally emerging residual asphericity in the NSC (which
effects were first studied in the context of resonant relaxation \citep{rau+96})
producing torques on the disk stars. Since the direction of the global
torque evolves randomly in time, the torque direction evolves, and
in turn, the disk orientation changes with time, leading to a flapping
like behavior, and warping of the disk (see Figure 2). It should be
noted that \citet{Kos+11} studied a resonant relaxation origin of
warping, though without self-consistent and realistic treatment of
the NSC; in particular they did not follow individual particles, and
therefore could not observe the important processes described in the
following. 

Besides the warping, a different process can be seen. The disk shows
a recurrent growth and destruction of a global $m=1$ unstable mode,
as it manifests itself in a single spiral arm structure in the disk,
thereby giving rise to large scale asymmetries in the distribution
of stars in the disk and significant over-densities (clumping) of
the stars. 

\begin{figure*}
\includegraphics{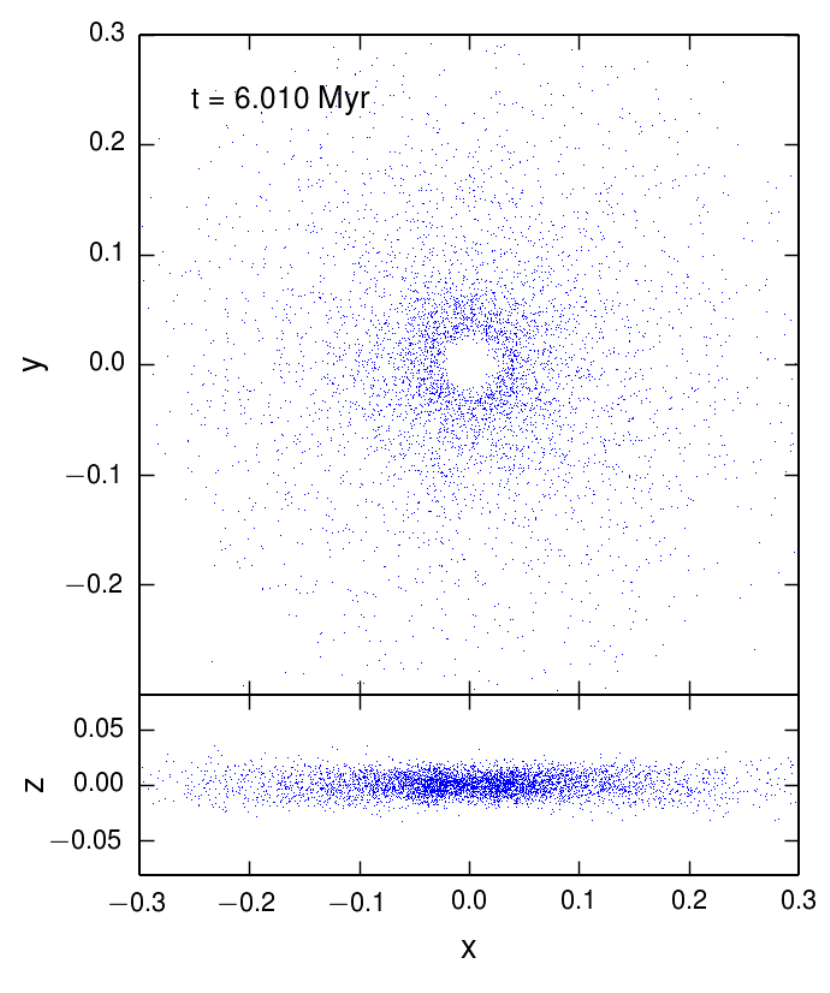}\includegraphics{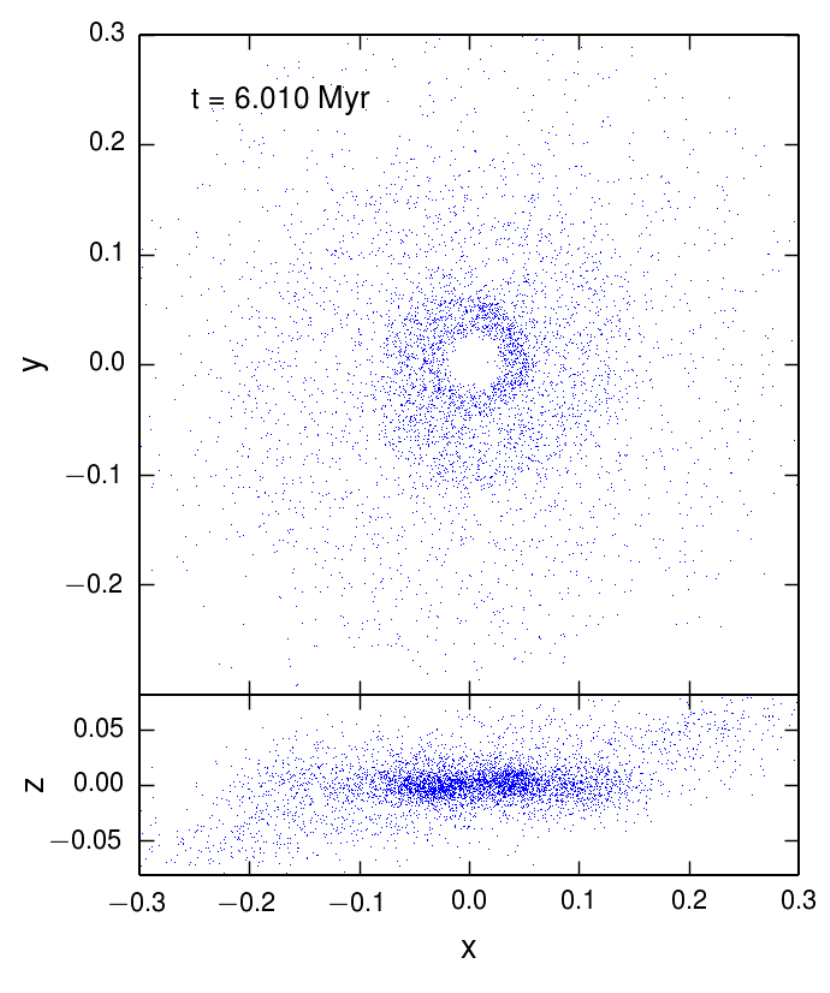}
\caption{\label{fig:6Myr-disk}The structure of a nuclear stellar disk after
6 Myrs of evolution (distance shown in pc). Left: A disk evolved in
isolation with no NSC, showing a smooth symmetric structure. Right:
A disk evolved with an NSC composed of live SBHs and a smooth potential
for the contribution of regular stars (model 7). In this model, a
significant spiral arm formed already at 1.5 Myrs, dissipated and
then reformed at $\sim5.5$ Myrs, and the disk changed its orientation
over time and became warped. }
\end{figure*}

\section{Discussion }
The spiral arm and asymmetry produced in the disk, together with warping
may help explain the origin of the puzzling properties of the nuclear
stellar disk in the Galactic center (see Figures 1-2). In particular,
these processes give rise to the possible existence of a spiral arm
structure and/or asymmetries and correlated clumping to be searched
for in current and future data. 

We suggest that the disk evolution is driven by the development of
a difference between the position of the central mass component (such
as the MBH in our case) and the center of mass of the system (such
as the NSC in our case), even irrespective of the disk. Such effects
were shown to play a role in the formation of $m=1$ unstable modes
in other astrophysical systems of different scales (such as galactic
disks on large scales, and protoplanetary systems on small scales;
see \citet{Jog+09} for an overview and references therein), but they
have been only little explored in the context of NSCs \citep{Tre05}.
Such processes can not be captured by considering a smooth external
potential, explaining the different evolution in models where the
NSC were only realized through such potentials. 

In the system explored here, the randomly changing potential from
the cusp SBHs and stars provides a changing environment. Such changes
may explain the recurrent formation and destruction of the spiral
structures and the random-walk-like behavior of the disk inclination,
compared with the coherent process occurring in the galactic-scale
and protoplanetary scale systems. Indeed, our simulations further
verify and pinpoint the collective nature of the process in even more
realistic models which account for all stars in the cusp. Our hybrid
modeling (models 8-12) realistically represents not only the live
SBHs, but also allows us to realistically model the contribution of
a live NSC, which computational cost in regular $N$-body simulations
would have been too prohibitive. The reproduction of the same behavior
in this case confirms that it is not the result of the more artificial
conditions used in the $N$-body simulations, and further supports
the robust nature of our results. Moreover, besides realizing a large
number of stars, the SCF modeling allows us to remove any effects
due to two-body relaxation, and consider only collective effects due
to the randomly induced overall change in the NSC potential over time.
In all cases where a discrete NSC is introduced we find a qualitatively
similar behavior, namely the recurrent formation and dissipation of
spiral arms, accompanied by a random walk of the disk inclination
and the development of disk warping. Such behavior is seen even when
the disk stars are treated as mass-less test particles (model 4),
i.e. the self-gravity of the disk does not play a major role in these
processes. 

Increasing the SBHs mass (e.g. taking $40$ $\mathrm{M}_{\odot}$
instead of $10$ ${\rm M_{\odot}}$, e.g. motivated by \citealt{Aha+16})
increases the randomly produced aspherical component of the NSC and
leads to a more pronounced and rapidly varying spiral arm structure.
Nevertheless, though the role of the SBHs is important, the same qualitatively-similar
behavior (spiral arms, warping and clumping) could be seen even in
a model in which only the stellar (live) component of NSC was considered
and no SBHs were included (model 12).

Currently, only the few tens brightest OB stars can be directly observed
in the GC stellar disk. Unfortunately, identifying a spiral arm structure
in such sparse data is challenging. Nevertheless, the $m=1$ mode
pattern gives rise to a global non-trivial structure that is reflected
by over-densities and asymmetries expected to exist in the disk. Figures
1-2 show the resulting structure of a disk during and after 6 Myrs
of evolution; for comparison we show similar data for the outcome
of a disk evolved in isolation (Figure 1, left), showing a smoother,
thinner disk with no evidence of asymmetries, nor warping nor clumping
as seen in the NSC-embedded disk. Consequently, even if a bona-fide
spiral structure could not be identified with current data, over-density
clumps are expected to exist. These may explain the
origin of the puzzling IRS13 clump without invoking the existence
of a (non-detected) IMBH. Moreover, our models predict the existence
of a clear asymmetry in the disk. Identifying the latter could be
currently difficult to disentangle from potential differential extinction
over different regions in the GC disk \citep{gen+10}, but could potentially
be better resolved in the future. 

\section{summary}
We explored the dynamics of the stellar disk orbiting the massive
black hole in the Galactic center through N-body and hybrid self-consistent-field
method simulations, accounting for the realistic effects of the stellar
cluster in which the disk is embedded. Our results reveal a novel
behavior of the nuclear stellar disk, producing spiral arms, warping,
assymetreis and clumping in the disk. We find that the collective
effects introduced by the nuclear stellar cluster play a key role
in determining the evolution of the disk and its structure. Such behavior
is robust, and can arise from a wide range of initial conditions.
Our results potentially provide an explanation for the origin of the
peculiar kinematic features in the observed GC disk (warping, clumping
and asymmetry), and give rise to additional predictions such as the
possible existence of a spiral arm structure which could be explored
with observational data in the coming years. Moreover, such processes
should similarly affect the evolution of other galactic nuclear disks,
including stellar disks formed in disks of active galactic nuclei
and may similarly affect the evolution of \emph{gaseous} disks in
galactic nuclei. Consequently, the same processes could be important
for the evolution of nuclear maser disks and active galactic nuclei,
and possibly even for their fueling at small, pc scales, in similar
ways as spiral structures in a large-scale galactic disk play a role
in the heating of the disk and fueling gas and stars into its inner
regions. 

\bibliographystyle{apj}

\end{document}